\documentclass[aps,twocolumn,prl,superscriptaddress]{revtex4-1}
\usepackage{epsfig,amssymb,amsmath,mathrsfs,amsthm,graphicx,float,color}

\def\Tr{\mathop{\rm Tr}\nolimits}
\def\d{\mathop{\rm d}\nolimits}

\newcommand{\map}[1]{\mathcal{#1}}

\newcommand{\set}[1]{\mathsf{#1}}
\newcommand{\grp}[1]{\mathsf{#1}}
\newcommand{\spc}[1]{\mathcal{#1}}

\def\>{\rangle}
\def\<{\langle}

{\catcode`\|=\active
  \gdef\Braket#1{\begingroup
\mathcode`\|32768\let|\BraVert\left<{#1}\right>\endgroup}}
\def\BraVert{\egroup\,\mid\,\bgroup}

\newtheorem{thm}{Theorem}
\newtheorem{lemma}{Lemma}

\definecolor{kmblue}{rgb}{0.19, 0.25, 0.91}
\definecolor{kmred}{rgb}{0.79, 0.29, 0.0}
\definecolor{kmgreen}{rgb}{0, 0.42, 0.24}

\begin{document}

\title{Optimal Compression for Identically Prepared Qubit States}
\author{Yuxiang Yang}
\affiliation{Department of Computer Science, The University of Hong Kong, Pokfulam Road, Hong Kong}
\author{Giulio Chiribella}
\affiliation{Department of Computer Science, The University of Hong Kong, Pokfulam Road, Hong Kong}
\affiliation{Canadian Institute for Advanced Research,
CIFAR Program in Quantum Information Science, Toronto, Ontario, M5G 1Z8, Canada}
\author{Masahito Hayashi}
\affiliation{Graduate School of Mathematics, Nagoya University, Nagoya, Japan}
\affiliation{Centre for Quantum Technologies, National University of Singapore, Singapore}

\begin{abstract}
We  establish  the ultimate limits to  the compression of  sequences of identically prepared qubits.   
 The  limits   are determined   by  Holevo's information quantity  and 
 are  attained  through use of  the optimal universal cloning machine,  which finds here a novel application to quantum Shannon theory. 
\end{abstract}

\maketitle

\noindent{\it Introduction.} A fundamental feature distinguishing quantum states from classical probability distributions is the freedom in the choice of basis, which can be used to encode information even when the spectrum of the state is fixed.   States  with fixed spectrum  can be used, for instance, as indicators of  spatial directions \cite{bagan-direction,chiribella-direction}, probes for frequency estimation \cite{mixed-clock1,mixed-clock2}  or even pieces of cryptocurrency \cite{
cirac}.  
Because of Holevo's bound \cite{holevo-1973}, the  basis information  cannot be extracted from  a single quantum particle, but   becomes accessible when multiple copies of the same quantum state are available.  
Suppose that  a sender  wants to transmit to a receiver the information contained in a sequence of $n$ identically prepared particles.    In this scenario, an important  question is how to minimize the amount of quantum bits (qubits) used in the transmission, subject to the requirement  that the initial  $n$-particle state can be approximately rebuilt  at the receiver's end.

The compression of  identically prepared states has been theoretically studied \cite{buzek} 
 and experimentally implemented \cite{rozema} in the pure state case. 
For mixed states, two of us proposed a protocol \cite{yang-chiribella-2016-prl} that  compresses states with fixed spectrum and variable  basis. The protocol  encodes $n$ identically prepared qubits into  a memory of $3/2 \log n$ qubits, which is proven to be the  smallest memory size  when the decoder is bound by the conservation of  the  total angular momentum.     Whether lifting the angular momentum constraint   allows for further compression  has remained an open problem so far. Moreover, little is known 
 in the case  where no prior information   is available on  the spectrum.  Finding the optimal compression protocol for  general quantum states is important for applications (where the spectrum may be unknown) and  for the  foundations of quantum theory, because it provides a  characterization of  
  the different information content of  quantum states and  classical probability distributions.

In this Letter we identify the optimal compression protocols for sequences of identically prepared qubits. 
We first consider states with known spectrum,   devising  a compression protocol that stores a sequence of $n$ qubits  into a memory of  $\log n$ qubits, the ultimate  limit set by  Holevo's  information  quantity \cite{holevo-1973}.   The memory reduction from $3/2 \log n$ to $\log n$ qubits is accomplished through a novel application of the optimal universal cloning machine  \cite{buzek-hillery,gisin-massar,werner}, here used to     modulate the values  of the total  angular momentum.  
On average, the modulation is  of size $\sqrt n$ and its logarithm is exactly  the amount of memory saved by our protocol, compared  to  the optimal protocol with angular momentum preserving decoder \cite{yang-chiribella-2016-prl}.  We then address a new compression scenario where no prior information about the  state is given. 
For this scenario, called {\em full-model compression}, we devise a protocol that uses a hybrid memory of $\log n$ qubits and $ 1/2 \log n$ classical bits. 
   The protocol is optimal; in fact,  no further compression can be achieved even if  the hybrid memory  is replaced by a  fully quantum memory.    
    The main result of the Letter is summarized by the following theorem:
\begin{thm}
A sequence of $n$ identically prepared qubit states  can be optimally compressed into $\log n$   qubits if the spectrum is known and into $\log n$ qubits plus $1/2 \log n$ classical bits  if the spectrum is  unknown. 
\end{thm}
Comparing the two protocols, we identify $\log n$ qubits as the amount of  information contained in the choice of basis and $1/2 \log n$ bits as the information contained in the spectrum.  This interpretation is consistent with the fact  that   $1/2 \log n$ is  the number of bits needed to faithfully compress  $n$ independent samples of a classical probability distribution over the binary set $\{0,1\}$ \cite{clarke-barron}. 


\noindent{\it Compression protocol for known spectrum.}  Consider the compression of $n$  qubits, independently prepared in the state  $\rho_g  =  g \rho  g^\dag$, where   $\rho  =  p \,  |0\>\<0|  +  (1-p)\,  |1\>\<1|$ is a fixed density matrix  and $g  \in  \grp{SU} (2)$ is a variable unitary matrix implementing a change of basis.    Without loss of generality, we assume $p\ge 1/2$ (the case $p<1/2$ is automatically accounted for by the change of basis).  
   Using  the Schur-Weyl duality \cite{fulton-harris}, the state of  the  $n$ qubits can be written  in  the block diagonal form
\begin{align}\label{decomp}
\rho_g^{\otimes n} = \bigoplus_{J=0}^{n/2}q_{J}\left(\rho_{g,J}\otimes\frac{I_{m_J}}{m_J}\right)  \, ,
\end{align}
where  the equality holds  up to a global unitary transformation,  known as  the Schur transform and efficiently implementable on a quantum computer 
\cite{bacon-chuang-2006-prl}. In Eq. (\ref{decomp}), $J$ is the quantum number of the total angular momentum \cite{footnote}, $q_J$ is a probability distribution, $\rho_{g,J}$ is a  density matrix  with support in an irreducible space   $\spc R_J$, and  $I_{m_J}$ is the identity matrix on an $m_J$-dimensional multiplicity space  $\spc M_J$ \cite{fulton-harris}.   The state $\rho_{g,J}$ can be expressed in the Gibbs  form \cite{purification}
\begin{align}
\nonumber \rho_{g,J} & =    \frac{e^{-\beta  H_{g,J}}}{\Tr\left[ e^{-\beta  H_{g,J}}\right]} \, ,  \qquad  \beta=  2  \tanh^{-1}  (2p-1)   \\
 \label{gibbs}   H_{g,J}& =   U_{g,J}  \, \left(  \sum_{m=-J}^J   -  m ~    |J,m\>\<J,m|  \right) U_{g,J}^\dag\, ,
\end{align} 
where  $\{|J,m\> \}_{m=  -J}^J$ are the eigenstates of the $z$ component of the  angular momentum operator  and $U_{g,J}$ is  the unitary matrix representing the change of basis $g$ in the irreducible space $\spc R_J$. 

We now show how to optimally compress the states $\rho_g^{\otimes n}$.  In general, a compression protocol consists of two components: the encoder, which stores the input state  into a memory, and the decoder, which  attempts to reconstruct the input state   from the state of the memory. The encoder and the decoder are both represented by  completely positive trace preserving linear maps (also known as quantum channels) \cite{structure}. Therefore, a quantum compression protocol  is  specified by a couple $(\map{E},\map{D})$, consisting of the encoding  and the decoding channel, respectively. The performance of the protocol is determined  by the tradeoff between two quantities: the memory size, quantified by  the dimension  $d_{\rm enc}$ of  the memory's Hilbert space, and 
 the compression error, measured by  the worst-case trace distance between the initial state and  the state recovered from the memory
  \begin{align}\label{error}
\epsilon&=\max_{g\in\grp{SU}(2)} \,  \frac12\left\|\map{D}\circ\map{E}\left(\rho_g^{\otimes n}\right)-\rho_g^{\otimes n}\right\|_1 \, ,
\end{align}
with $\|  A\|_1 :  =  \Tr \sqrt {A^\dag A}$. The key issue is to minimize the memory size, while guaranteeing that the compression error   vanishes  in the large  $n$ limit.

The optimal protocol is based on two ingredients:  The first is the concentration of the probability distribution  $q_J$ in Eq. (\ref{decomp}).  Explicitly, the probability is given by    \cite{yang-chiribella-2016-prl} 
\begin{align}
q_{J}=\frac{2J+1}{2J_0}&\left[  B\left(\frac n2 +  J+1 \right) 
-B\left(\frac n2 - J \right)\right]\label{qJ}
\end{align}
where   $B(k)$ is the binomial distribution with $n+1$ trials and probability $p$ and    $J_0 := (p-1/2)(n+1)$ is close to the average value $\<J\>   =  \sum_J  \,  J \, q_J$. From the above expression  it is clear  that   
the values of $J$ with $|J-J_0|  \gg \sqrt n$ have exponentially small probability in the large $n$ limit. As a result, the performance of a compression protocol depends only on its action on the subspaces $\spc R_J\otimes \spc M_J$  that satisfy the condition  $|J-J_0|  =  O(\sqrt n)$.  

The second ingredient of our compression protocol is a remarkable property of the optimal universal cloning machine (UCM) \cite{gisin-massar,werner}.  Mathematically, the UCM  is described by a map  transforming (operators supported in) the symmetric subspace of $2J$ qubits into (operators supported in) the symmetric subspace of $2K$ qubits.     Here we allow $J$ to be larger than $K$, in which case the ``cloning" process   just consists in getting rid of  $2(J-  K)$ qubits.  
  With this convention, the cloning channel is 
\begin{align}\label{element-channel}
\map{C}_{J\to K}(\rho)=\left\{\begin{array}{ll} \left(\frac{2J+1}{2K+1}\right)   P_{K}\left(\rho\otimes   P_{K-J}
\right)P_{K} & J\le K\\ \Tr_{2(J-K)}[\rho]  & J>K \end{array}\right.  
\end{align}
where $P_{x}$  is the projector  on the symmetric subspace of $2x$ qubits  and $\Tr_x$ denotes the partial trace over the first $x$ qubits.  
The key to our compression protocol is to regard the Gibbs  states in Eq. (\ref{gibbs})  as states on the symmetric subspace of $2J$ qubits and to observe that UCM has the following  property, derived in the Appendix:
\begin{lemma}\label{thm-info-block}
The universal cloning channel $\map{C}_{J\to K}$   
 transforms the Gibbs state $\rho_{g,J}$ into the Gibbs state $\rho_{g,K}$ with  error
 \begin{align}\label{e-bound}
  \left\|      \map C_{J\to K}  \left( \rho_{g,J}\right)  -   \rho_{g,K}  \right\|_1 \le \delta^{1-s}+O\left(\delta \right) \, ,  
\end{align}
where $s>0$ is an arbitrary constant and $\delta:={{|J-K|}/{J}}$.
\end{lemma}
\begin{figure}[t!]
      \includegraphics[width=0.45\textwidth]{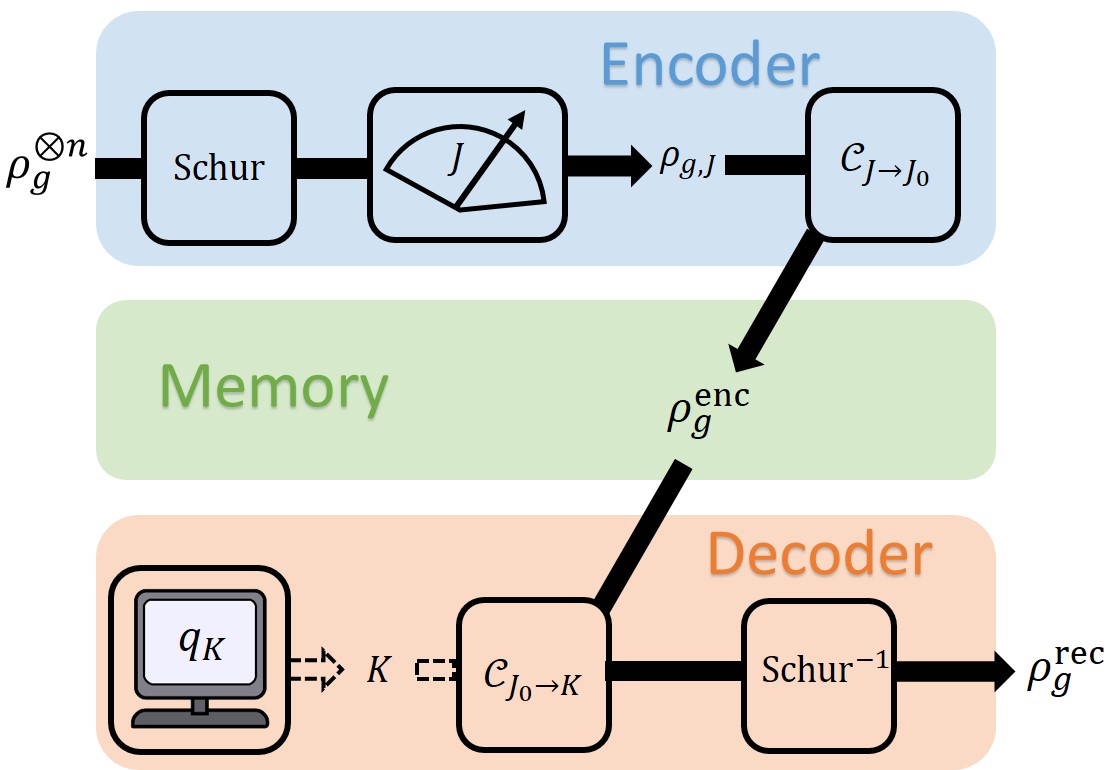}\caption{{\bf Optimal compression for known spectrum and completely unknown basis.} The encoder collects information from  subspaces with different angular momenta and concentrates it into a system with angular momentum  $J_0 $. The decoder spreads the information back,  modulating the angular momentum by   $\sqrt n$ units on average. 
      }
       \label{fig:known}
\end{figure}
This result establishes  a bridge between the  cloning of pure states and the compression of  mixed  states.    
 Leveraging on Lemma \ref{thm-info-block} and on the concentration of the probability distribution $\{q_J\}$,  we  devise  the following protocol: 
\begin{itemize}
\item {\it Encoder.} Perform the Schur transform.  Then, measure the quantum number $J$ with the nondemolition measurement that preserves the quantum information in each subspace   $\map R_J\otimes \map M_J$.   Discard  the multiplicity register  and  apply the cloning channel  $\map{C}_{J\to J_0}$ 
  to the remaining state $\rho_{g,J}$.  Store  the output state  $\map{C}_{J\to J_0}(\rho_{g,J})$ 
   into a quantum memory of dimension $d_{\rm enc}  =    2J_0 +1$.  

\item{\it Decoder.} Pick a value $K$ at random with probability $q_K$ and apply the cloning channel $\map{C}_{J_0\to   K}$ to the quantum memory.    Append a multiplicity register in the maximally mixed state $I_{m_{K}}/m_{K}$.  Finally,  perform the inverse of the Schur transform. 
\end{itemize}
The protocol,  illustrated in Fig. \ref{fig:known}, is mathematically described by the channels
   \begin{align}
\nonumber \map{E}(\rho)  &=\sum_{J=0}^{n/2} \map{C}_{J\to J_0}
\left[\Tr_{\spc{M}_J}\left(\Pi_J \rho \Pi_J \right) \right] \\
\map{D}(\rho)&=\bigoplus_{K=0}^{n/2} q_K \left[ \map{C}_{J_0\to K}(\rho)\otimes\frac{I_{m_{K}}}{m_{K}}\right] \, ,
\end{align}
where $\Pi_J$ is the projector on $\spc R_J\otimes \spc M_J$ and $\Tr_{\spc{M}_J}$ denotes the partial trace over $\spc M_J$.  

The above protocol requires a memory of $ \log (2J_0 +1)  =  \log n  +  O(1)$ qubits. 
On the other hand, the error is arbitrarily small for large $n$:   this is because   the states $\rho_{g,J}$ with $|J -  J_0|\gg \sqrt n$ have negligible probability according to Eq. (\ref{qJ}),  while the states $\rho_{g,J}$  with  $|J -  J_0|  =  O( \sqrt n)$ can be faithfully encoded in the state $\rho_{g,J_0}$,  thanks to Lemma \ref{thm-info-block} (see  the Appendix for more details). 
  
\noindent{\it Optimality of the protocol with known spectrum.}
 Our protocol uses the minimum memory size compatible with the requirement of vanishing error.  The argument goes as follows:  For a generic ensemble  $\set{E}=\{\rho_x, p_x\}$,  a measure of the information content  is provided by Holevo's  information \cite{holevo-1973}   
  \begin{align}\label{chi}
\chi\left(   \set E \right)=H\left(\sum_{x}p_x\rho_x\right)-\sum_x p_x H(\rho_x) 
\end{align}
where $H  (\rho)  =  -\Tr [\rho \log \rho]$  is  the  von Neumann entropy. When the ensemble $\set E$ is  faithfully stored in a quantum memory, the memory should be large enough to accommodate the Holevo information of $\set E$.  Since   a memory of dimension $d_{\rm enc}$  can have at most   a Holevo information of $\log d_{\rm enc}$   \cite{holevo-1973}, one has the bound $\log d_{\rm enc} \ge \chi(\set E)$.  For $\epsilon >0$, an approximate version of the bound is \cite{wilde}
 \begin{align}
\log d_{\rm enc} &\ge\chi\left(\set{E}\right)-2  \epsilon\log d_{\set{E}}-2 \mu(\epsilon)  \, ,\label{Holevo}
\end{align}
  where $d_{\set{E}}$   is the effective dimension, defined as  the rank of the average state $\rho_{\set E}  :=  \sum_x p_x\rho_x$, and    $\mu(\epsilon):=-\epsilon\ln \epsilon$. 


Equation (\ref{Holevo})  sets a lower bound  on the memory size,  valid for arbitrary ensembles.   However,   the bound may not be tight. Notably, the bound  is \emph{not} tight  for the ensembles considered in our paper.  
  The reason is the dimension-dependent  term $\log d_{\set E}$, which can be arbitrarily large: in our case, we have $d_{\set E}  =  2^n$ for $p\not  =   0,1$.  To address this problem,   we use  the notion of       sufficient statistics   \cite{petz-1986-cmp}.  
   An ensemble $\set E'   =  \{\rho_x'  ,  p_x\}$  is called a \emph{sufficient statistics for the ensemble $\set E  = \{\rho_x, p_x\}$}  if the states of $\set E$ can be  encoded into states of $\set E'$ and decoded with zero  error. 
        Since the encoding is reversible,  the ensembles $\set E$ and $\set E'$ have the same  Holevo information, namely  $\chi  (\set E')   =  \chi (\set E)$. Moreover, the number of qubits needed to encode the original ensemble $\set{E}$ up to error $\epsilon$ is equal to the number of qubits needed to encode the ensemble $\set E'$, up to the same error (see the Appendix for more detail).    Using these facts, we can improve the bound (\ref{Holevo}), obtaining
   \begin{align}
\log d_{\rm enc} &\ge\chi\left(\set{E}\right)-2  \epsilon\log d^{\min}_{\set{E}}-2 \mu(\epsilon)  \, ,\label{Holevo+}
\end{align}     
where $d_{\set E}^{\min}$ is the minimum of $d_{\set E'}$ over all ensembles $\set E'$ that are    sufficient statistics for $\set E$.  We call Eq. (\ref{Holevo+}) the \emph{Holevo  bound for compression}.

Let us  apply the bound to the ensemble     $\set E  =  \{\rho_g^{\otimes n} ,    \d g\}$, where  $\d g$ represents  the uniform  distribution over all changes of basis.   For this ensemble,  explicit calculation (provided in the Appendix)  yields  
\begin{align}\label{chi1}
\chi (\set E)& =   \log n+O(1) \,  .
\end{align}
 A sufficient statistics for $\set E$ is provided by the ensemble  $\set E'  =  \{  \rho_g'  ,  \,  \d g  \}$ with 
$\rho_g'   :  =  \bigoplus_{J=0}^{n/2} q_J  \,    \rho_{g,J}$, obtained by getting rid of the multiplicity spaces in Eq. (\ref{decomp}). 
The ensemble $\set E'$   has effective   dimension  
\begin{align}\label{dsuff}
d_{\set{E}'}=\sum_{J=0}^{n/2}  \, (2J+1)   = \left( \frac n 2+1\right)^2 \,  ,
\end{align}  
which has been proven to be the minimum over all  sufficient statistics     \cite{koashi-imoto-2001-prl,yang-chiribella-2016-prl}.  Inserting Eqs. (\ref{chi1}) and (\ref{dsuff}) into Eq. (\ref{Holevo+}) we  obtain the bound
\begin{align}\label{bound-final}
\log d_{\rm enc} &\ge   (1-4\epsilon) \, \log n   +4\epsilon-2\mu(\epsilon)+O(1) \, .
\end{align}
When $\epsilon$ is asymptotically small, the leading  term is  $\log n$, the number of qubits used by  our protocol. Hence, we conclude that the protocol is optimal and that the Holevo bound for compression is tight for the ensemble $\set E$.

\begin{figure}[t!]
      \includegraphics[width=0.45\textwidth]{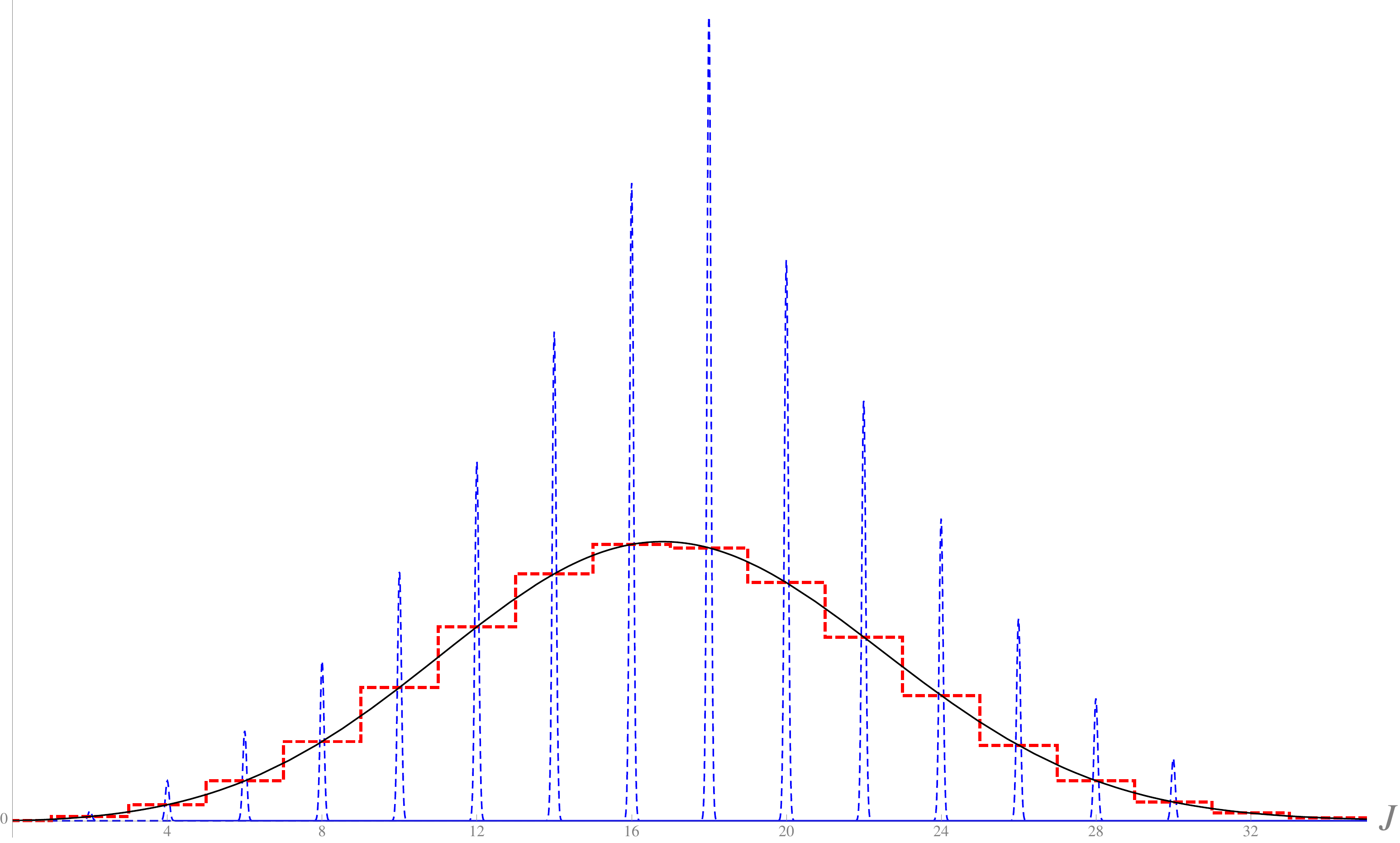}\caption{{\bf Spectral distributions of the output states with and without sampling.} A comparison of the spectral distributions of the following states: the original state $\rho_g^{\otimes n}$ (black, solid line), the output state of the optimal protocol (red, dashed line), and the output state of a protocol with the same encoder of the optimal protocol and a decoder   without sampling (blue, dashed line). }
       \label{fig:compare}
\end{figure}

\noindent{\it Compression protocol for arbitrary qubit states.}  Let us now turn to the full-model compression. 
  A simple protocol for compressing arbitrary states is  to measure the magnitude of the total angular momentum, to store the outcome $J$ in a classical memory and the state  $\rho_{g,J}$ in a quantum memory.  Since  $J$ can take any value between 0 and $n/2$,  this protocol requires  $\lceil \log   ( n/2+1  )\rceil$ classical bits. Moreover, since $\rho_{g,J}$ has support in a $(2J+1)$-dimensional space, the protocol  requires  $  \lceil \log  (n+1) \rceil$ qubits in the worst case scenario.   At first sight,   it seems difficult to do any better: One cannot use less than $\log n$ qubits, because the input state could   consist of $n$ copies of a random pure state and no protocol can  compress such a state in  less than $\log n$ qubits \cite{yang-chiribella-2016-prl}. On the other hand,   $J$ can take $n/2+1$ values and it is not possible to encode  this information in less than $\log n$  bits.     Despite these facts, we now show that   the amount of classical bits can be cut down  by half  with asymptotically negligible error.
The key idea is that the decoder  need not have  full  information about $J$: thanks to Lemma 
\ref{thm-info-block}, 
two states $\rho_{g,J}$ and $\rho_{g,K}$ with $|J-K|   =  O(\sqrt n)$ are approximately interconvertible.  Motivated by this fact, we partition the  values of $J$  into   disjoint intervals  $\set{L}_1,  \dots ,\set{L}_{t}$  of size $O(\sqrt n)$. Instead of encoding the measurement outcome $J$, we compute the index $i$ such that $J\in\set{L}_{i}$ and store it in a classical memory. Since the index $i$ can take  $O(\sqrt{n})$ values, the size of the  memory   is  $(1/2)\log n$, instead of $\log n$.  
  The details of the protocol are as follows: \begin{figure}[t!]
      \includegraphics[width=0.5\textwidth]{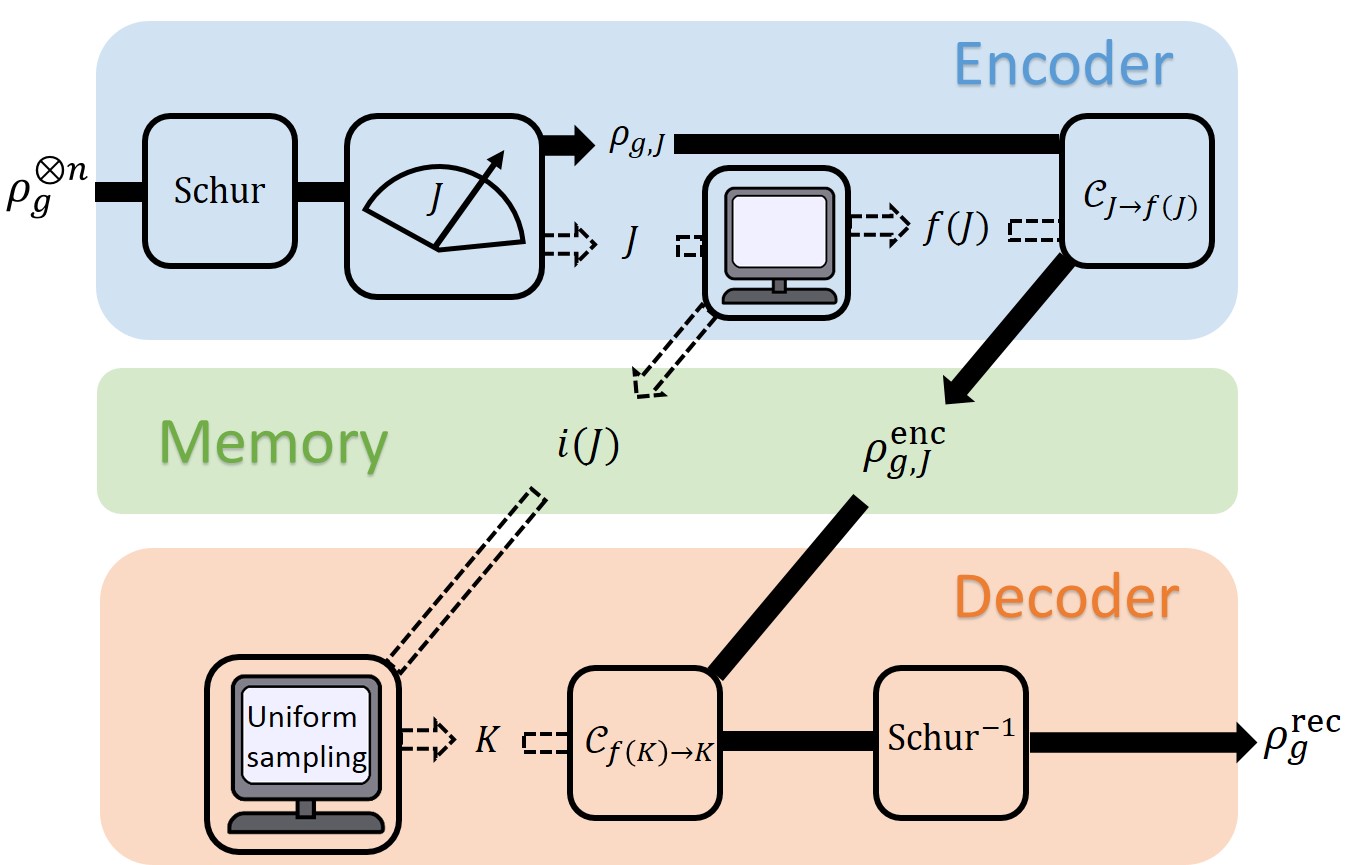}\caption{{\bf Optimal full-model compression.}   The encoder disassembles an arbitrary sequence of $n$ identically prepared qubits into a classical part ($1/2 \log n$ bits) and a quantum part ($\log n$ qubits).  The decoder recombines these two pieces of information, approximately retrieving the initial state of the sequence.
      }
       \label{fig:unknown}
\end{figure}
\begin{itemize}
\item {\it Encoder.} Perform the Schur transform.   Then,  measure the quantum number $J$  with the nondemolition measurement that preserves the quantum information in each   subspace  $\spc R_J\otimes \spc M_J$. Find the index $i (J)$ such that $J \in  \set L_{i(J)}$.     Discard the multiplicity register and   send the remaining state $\rho_{g,J}$ to the input of  the quantum channel $\map{C}_{J\to f(J)}$, where $f(J)$ is  the median of the subset $\set{L}_{i(J)}$.  Store the  output state $\map{C}_{J\to f(J)}(\rho_{g,J})$ in a quantum memory and the  index $i (J)$ in a classical memory.  

\item{\it Decoder.} 
Read the value of $i (J)$ from the classical memory.  For a given value of $i (J)$, pick  a random value $K$  in the subset $\set{L}_{i (J)}$ and  apply  the  channel $\map{C}_{f(K)\to   K}$ to the quantum memory.      Then, append the multiplicity register in the maximally mixed state  $I_{m_K}/m_K$.  Finally,  perform the inverse of the Schur transform. 
\end{itemize}
The  protocol is illustrated in Fig. \ref{fig:unknown}. The explicit expression of the channels $\map E$ and $\map D$, as well as the proof that the error vanishes in the large $n$ limit can be found in the Appendix. Here we emphasize a few points:
First,  it is convenient to choose one interval---say,  $\set L_t$---to contain only the value $J=  n/2$. In this way,  the protocol acts as the identity in the symmetric subspace and  pure states are compressed without error.  
Second, random sampling in the decoder is essential for achieving  vanishing error. This fact is illustrated in Fig. \ref{fig:compare},  which shows that  sampling yields a well-behaved interpolation of the spectral distribution in Eq. (\ref{qJ}), while the lack of sampling leads to a  poor approximation.   Third,  comparing the full model compression with the fixed-spectrum compression  leads us to identify $1/2\log n$ bits as the amount of  memory needed to store the  information about the spectrum. This interpretation is consistent with the fact that $1/2\log n$ bits is the size of the smallest classical memory needed to  faithfully store $n$  samples of a generic probability distribution over the set $\{0,1\}$ \cite{clarke-barron}.

\noindent{\it Optimality for the full-model compression:} The optimality of the full-model protocol can be proven with the same techniques used for fixed spectrum.  In fact, an even stronger result holds: replacing the hybrid memory with a fully quantum memory does not improve  the compression, because  $3/2 \log n$  qubits is the minimum memory size   allowed by the Holevo bound for compression. The details are provided in the Appendix.

\noindent{\it Conclusion:} In this Letter we showed how to compress identically prepared qubits in the smallest possible memory.  
 The key technique is the use of universal cloning to convert Gibbs states of different angular momentum.  Converting Gibbs states is  a novel application of quantum cloning  \cite{review-cloning, cerf, fan} and  may inspire further applications in the resource theory of quantum thermodynamics, both in the free  \cite{brandao-2013-prl} and in the size-restricted case \cite{tajima-hayashi}. 
   Extending our results, it is also interesting to investigate the relation between cloning and compression for  other families of states, such as phase  \cite{covariant,
broadcasting1} and mirror-phase \cite{mirror}  covariant states, and mixed states of arbitrary  finite dimensional systems \cite{yang-chiribella-2016-prl}. 
    The recent   implementations of various quantum cloning machines  \cite{
exp-clone2, exp-clone3,exp-clone4,exp-clone5}  suggests that   prototypes of optimal compression  may be experimentally demonstrated in the near future. 


\begin{acknowledgments}
We acknowledge the referees of this Letter for useful
suggestions that helped improve the presentation. G. C. is
supported by the Canadian Institute for Advanced Research
(CIFAR), by the Hong Kong Research Grant Council
through Grant No. 17326616, by National Science
Foundation of China through Grant No. 11675136, and
by the HKU Seed Funding for Basic Research. Y. Y. is
supported by a Hong Kong and China Gas Scholarship.
M. H. is partially supported by a MEXT Grant-in-Aid for
Scientific Research (A) No. 23246071 and the Okawa
Research Grant. Centre for Quantum Technologies is a
Research Centre of Excellence funded by the Ministry of
Education and the National Research Foundation of
Singapore. This work was completed during the
``Hong Kong Workshop on Quantum Information and
Foundations,” organized with support from the
Foundational Question Institute (FQXi-MGA-1502).
\end{acknowledgments}


%

\begin{appendix}
\begin{widetext}

\section{ Proof of Lemma 1.}

In this section, we show that  the universal cloning channel $\map C_{J\to K}$ transforms the Gibbs state $\rho_{g,J}$ into an approximation of the Gibbs state $\rho_{g,K}$, which  becomes accurate when $|J-K|/J$ is small.   Specifically, we show that the error satisfies the bound
\begin{align}
\frac12\left\|\map{C}_{J\to K}(\rho_{g,J})-\rho_{g,K}\right\|_1  \le\frac{\delta^{1-s}}{2}[1+O(\delta^s )]  \, ,\qquad \delta:=\frac{|J-K|}{J}   \, ,
\end{align}
valid for arbitrary $g\in\grp{SU} (2)$ and arbitrary $s>0$. 

First of all, note that the covariance of the cloning channel  and the unitary invariance of the trace norm imply the equality  
\begin{align}
\left\|\map{C}_{J\to K}(\rho_{g,J})-\rho_{g,K}\right\|_1  =  \left\|\map{C}_{J\to K}(\rho_{e,J})-\rho_{e,K}\right\|_1  \qquad \forall g \in  \grp{SU} (2)  \, ,
\end{align}
where $e$ is the identity element in $\grp {SU} (2)$.  Hence, it is enough to show the bound  
\begin{align}
 \frac12\left\|\map{C}_{J\to K}(\rho_{J})-\rho_{K}\right\|_1 &\le\frac{\delta^{1-s}}2[1+O(\delta^s )] \, ,\label{lemma}
\end{align}
with $\rho_{J} : =  \rho_{e,J}$ and $\rho_{K} : =  \rho_{e,K}$.  To prove this bound, we use the  expansion
\begin{align}\label{rho-decomp}
\rho_{J}=(N_J)^{-1}\sum_{m=-J}^{J}p^{J+m}(1-p)^{J-m}|J,m\> \<J,m| \, ,
\end{align}
where   $N_J$ is the normalization constant given by
  \begin{align}
  \nonumber  N_J  &=\sum_{j=-J}^{J}p^{ J+j}(1-p)^{J-j}  \\
 \label{NJ}
   &  =  p^{2J+1} \,    \frac { 1  -  \left (  \frac {1-p}p \right)^{2J+1}}{2p-1} \, .
  \end{align} 
In the following we will analyze the cases $J  \le K$ and $J>K$ separately. 

\subsection{The $J\le K$ case.}
We begin by checking the action of $\map{C}_{J\to K}$ on the projectors  $|J,m\>\<J,m|$. For $J\le K$ we have 
\begin{align*}
\map{C}_{J\to K}(|J,m\>\<J,m|)&=\left(\frac{2J+1}{2K+1}\right)P_{K}(|J,m\>\<J,m|\otimes P_{K-J})P_{K}\\
&=\left(\frac{2J+1}{2K+1}\right)\sum_{k}{2K-2J\choose K-J+k-m}{2J\choose J+m}{2K\choose K+k}^{-1}|K,k\>\<K,k|  \, .
\end{align*}

Note that we have the equality  
\begin{align*}
\<K,K+m-J| \, \map{C}_{J\to K}(|J,m\> \, \<J,m| ) \, |K,K+m-J\>=\left(\frac{2J+1}{2K+1}\right)\frac{{2J\choose J-m}}{{2K\choose J-m}} .
\end{align*}
Therefore, we can express $\map{C}_{J\to K}(|J,m\>\<J,m|)$ as 
\begin{align*}
\map{C}_{J\to K}(|J,m\>\<J,m|)=\left(\frac{2J+1}{2K+1}\right)\frac{{2J\choose J-m}}{{2K\choose J-m}}|K,K+m-J\>\<K,K+m-J| +\left[1-\left(\frac{2J+1}{2K+1}\right)\frac{{2J\choose J-m}}{{2K\choose J-m}}\right]\sigma_{J,K} 
\end{align*}
where $\sigma_{J,K}$ is a suitable  quantum state. Combining the above equation with Eq. (\ref{rho-decomp}), we obtain 
\begin{align*}
\map{C}_{J\to K}(\rho_{J})&=\sum_{m=-J}^{J}\frac{p^{J+m}(1-p)^{J-m}}{N_J}\left\{\left(\frac{2J+1}{2K+1}\right)\frac{{2J\choose J-m}}{{2K\choose J-m}}|K,K+m-J\>\<K,K+m-J|\right.\\
&\qquad\left.+\left[1-\left(\frac{2J+1}{2K+1}\right)\frac{{2J\choose J-m}}{{2K\choose J-m}}\right]\sigma_{J,K} \right\}.
\end{align*}
Now, we focus on the entries with $m\in[J-\lfloor\delta^s\rfloor,J]$, where $s>0$ is a parameter to be specified later. We  rewrite the output state as
\begin{align*}
\map{C}_{J\to K}(\rho_{J})&=\sum_{m=J-\lfloor\delta^{-s}\rfloor}^{J}\frac{p^{J+m}(1-p)^{J-m}}{N_J}\left(\frac{2J+1}{2K+1}\right)\frac{{2J\choose J-m}}{{2K\choose J-m}}|K,K+m-J\>\<K,K+m-J| +\mu_{J,K}  \, ,
\end{align*}
where  $\mu_{J,K}$ is a positive operator with trace 
\[ \Tr[\mu_{J,K}]=1-\sum_{m=J-\lfloor\delta^{-s}\rfloor}^{J}\frac{p^{J+m}(1-p)^{J-m}}{N_J}\left(\frac{2J+1}{2K+1}\right)\frac{{2J\choose J-m}}{{2K\choose J-m}} \, .\]
 Next, substituting $J-m$ with $k$, we have
\begin{align*}
\map{C}_{J\to K}(\rho_{J})&=\sum_{k=0}^{\lfloor\delta^{-s}\rfloor}\frac{p^{2J-k}(1-p)^{k}}{N_J}\left(\frac{2J+1}{2K+1}\right)\frac{{2J\choose k}}{{2K\choose k}}|K,K-k\>\<K,K-k| +\mu_{J,K}.
\end{align*}
  Using the expression (\ref{rho-decomp}) for $\rho_K$,  we  bound the error as 
 \begin{align*}
 \frac12\|\map{C}_{J\to K}(\rho_{J})-\rho_{K}\|_1 &=\frac12\left\|\sum_{k=0}^{\lfloor \delta^{-s}\rfloor}(1-p)^k\left[\frac{p^{2J-k}}{N_J}\left(\frac{2J+1}{2K+1}\right)\frac{{2J\choose k}}{{2K\choose k}}-\frac{p^{2K-k}}{N_{K}}\right]|K,K-k\>\<K,K-k|+\mu_{J,K}\right.\nonumber\\
&\qquad\left.-\sum_{k=\lfloor\delta^{-s}\rfloor+1}^{2K}\frac{p^{2K-k}(1-p)^k}{N_{K}}|K,K-k\>\<K,K-k|\right\|_1\\
&\le \frac12\sum_{k=0}^{\lfloor\delta^{-s}\rfloor}(1-p)^kp^{-k}\left|\frac{p^{2J}}{N_J}\left(\frac{2J+1}{2K+1}\right)\frac{{2J\choose k}}{{2K\choose k}}-\frac{p^{2K}}{N_{K}}\right|+\frac12\Tr[\mu_{J,K}]+\frac12\sum_{k=\lfloor\delta^{-s}\rfloor+1}^{2K}\frac{p^{2K-k}(1-p)^k}{N_{K}}\\
&\le\frac{p}{4p-2}\max_{k\in[0,\lfloor\delta^{-s}\rfloor]}\left|\frac{p^{2J}}{N_J}\left(\frac{2J+1}{2K+1}\right)\frac{{2J\choose k}}{{2K\choose k}}-\frac{p^{2K}}{N_{K}}\right|+\frac12\Tr[\mu_{J,K}]+\left(\frac{1-p}{p}\right)^{\lfloor\delta^{-s}\rfloor}
\end{align*}
Since $p>1/2$ and $s>0$, it is obvious that the third term in the last inequality vanishes exponentially in $J$, and we need only to show that the first term and the second term also vanish as $J$ grows.

For the first error term, we have the following expansion:
\begin{align*}
&\left|\frac{p^{2J}}{N_J}\left(\frac{2J+1}{2K+1}\right)\frac{{2J\choose k}}{{2K\choose k}}-\frac{p^{2K}}{N_{K}}\right|\\
=&\frac{p^{2K}}{N_{K}}\left|p^{2J-2K}\left(\frac{2J+1}{2K+1}\right)\frac{N_{K}}{N_{J}}\frac{{2J\choose k}}{{2K\choose k}}-1\right|\\
=&\frac{p^{2K}}{N_{K}}\left|\left(\frac{2J+1}{2K+1}\right)\frac{1-\left(\frac{1-p}{p}\right)^{2K+1}}{1-\left(\frac{1-p}{p}\right)^{2J+1}}\frac{{2J\choose k}}{{2K\choose k}}-1\right|\\
=&\frac{p^{2K}}{N_{K}}\left|\left(\frac{2J+1}{2K+1}\right)\frac{1-\left(\frac{1-p}{p}\right)^{2K+1}}{1-\left(\frac{1-p}{p}\right)^{2J+1}}e^{k\ln\left(\frac{J}{K}\right)+(2K-k+1)\ln\left(1-\frac{k}{2K}\right)-(2J-k+1)\ln\left(1-\frac{k}{2J}\right)+O\left(\frac{1}{J}\right)}-1\right|  \, ,
\end{align*}
the third line coming from  Eq. (\ref{NJ}). 
Recalling that $\delta=(K-J)/J$, it is straightforward to verify that 
\begin{align*}
\frac{2J+1}{2K+1}&=1-\delta+O(\delta^{2})\\
e^{k\ln(J/K)}&=1-k\delta+O(k\delta^{2})\\
e^{(2K-k+1)\ln[1-k/(2K)]-(2J-k+1)\ln[1-k/(2J)]}&=1-\frac{k^2  \delta}{4K}+O\left(k^3\delta J^{-2}\right).
\end{align*}
Substituting the above equations into the expression of the first error term, we have
\begin{align*}
\max_{k\in[0,\lfloor\delta^{-s}\rfloor]}\left|\frac{p^{2J}}{N_J}\left(\frac{2J+1}{2K+1}\right)\frac{{2J\choose k}}{{2K\choose k}}-\frac{p^{2K}}{N_{K}}\right|&=\frac{p^{2K}}{N_{K}}\max_{k\in[0,\lfloor\delta^{-s}\rfloor]}\left|-\delta-k\delta-\frac{k^2  \delta}{4K}+O\left(k^3\delta J^{-2}\right)+O(k\delta^2)+O(J^{-1})\right|\\
&\le \frac{p^{2K}}{N_{K}}\left[\delta^{1-s} +O(\delta)+O(J^{-1})\right]\\
&=\frac{2p-1}{p}\cdot \delta^{1-s} [1+O(\delta^s )].
\end{align*}
For the second error term, we have
\begin{align*}
\Tr[\mu_{J,K}]&=1-(N_J)^{-1}\sum_{m=J-\lfloor \delta^{-s}\rfloor}^{J}p^{J+m}(1-p)^{J-m}\left(\frac{2J+1}{2K+1}\right)\frac{{2J\choose J-m}}{{2K\choose J-m}}\\
&\le 1-(N_J)^{-1}\sum_{m=J-\lfloor \delta^{-s}\rfloor}^{J}p^{J+m}(1-p)^{J-m}\left(\frac{2J+1}{2K+1}\right)\min_{m'\in[J-\lfloor \delta^{-s}\rfloor,J]}\frac{{2J\choose J-m'}}{{2K\choose J-m'}}\\
&\le 1-(N_J)^{-1}\sum_{m=J-\lfloor \delta^{-s}\rfloor}^{J}p^{J+m}(1-p)^{J-m}\left(\frac{2J+1}{2K+1}\right)\\
&=1-\frac{1-\left(\frac{1-p}p\right)^{\delta^{-s}+1}}{1-\left(\frac{1-p}p\right)^{2J+1}}\left(\frac{2J+1}{2K+1}\right)\\
&\le\delta,
\end{align*}
which vanishes as $J$ grows. Finally, combining the above calculations, the error of the conversion can be bounded as
\begin{align*}
 \frac12\|\map{C}_{J\to K}(\rho_{J})-\rho_{K}\|_1&\le\frac{\delta^{1-s}}2[1+O(\delta^s )],, 
 \end{align*}
for any $s>0$. Since $s$ can be chosen to be arbitrarily small, the leading order of the error is close to $\delta$. 

\subsection{The $J>K$ case.}
In this case, the action of $\map{C}_{J\to K}$ on the projectors $|J,m\>\<J,m|$  is
\begin{align*}
\map{C}_{J\to K}(|J,m\>\<J,m|)=\sum_k{2J-2K\choose J-K+m-k}{2K\choose K+k}{2J\choose J+m}^{-1}|K,k\>\<K,k| \, .
\end{align*}
Notice that 
\begin{align*}
\<K,K+m-J|\map{C}_{J\to K}(|J,m\>\<J,m| )|K,K+m-J\> =\frac{{2K\choose J-m}}{{2J\choose J-m}} .
\end{align*}
Therefore, we can express $\map{C}_{J\to K}(|J,m\>\<J,m|)$ as 
\begin{align*}
\map{C}_{J\to K}(|J,m\> \<J,m|)=\frac{{2K\choose J-m}}{{2J\choose J-m}}|K,K+m-J\>\<K,K+m-J|+\left[1-\frac{{2K\choose J-m}}{{2J\choose J-m}}\right]\sigma_{J,K} 
\end{align*}
where $\sigma_{J,K}$ is a  suitable quantum state. Combining the above equation with Eq. (\ref{rho-decomp}), we have
\begin{align*}
\map{C}_{J\to K}(\rho_{J})&=(N_J)^{-1}\sum_{m=-J}^{J}p^{J+m}(1-p)^{J-m}\left\{\frac{{2K\choose J-m}}{{2J\choose J-m}}|K,K+m-J\>\<K,K+m-J| +\left[1-\frac{{2K\choose J-m}}{{2J\choose J-m}}\right]\sigma_{J,K} \right\}.
\end{align*}
Again, we focus on the entries with $m\in[J-\lfloor \delta^{-s}\rfloor,J]$ for a parameter $s>0$ and rewrite the output state as
\begin{align*}
\map{C}_{J\to K}(\rho_{J})&=(N_J)^{-1}\sum_{m=J-\lfloor  \delta^{-s}\rfloor}^{J}p^{J+m}(1-p)^{J-m}\frac{{2K\choose J-m}}{{2J\choose J-m}}|K,K+m-J\>\<K,K+m-J|+\mu_{J,K}.
\end{align*}
Here $\mu_{J,K}$ is a positive operator with trace $\Tr[\mu_{J,K}]=1-(N_J)^{-1}\sum_{m=J-\lfloor \delta^{-s}\rfloor}^{J}p^{J+m}(1-p)^{J-m}\frac{{2K\choose J-m}}{{2J\choose J-m}}$. Next, substituting $J-m$ with $k$, we have
\begin{align*}
\map{C}_{J\to K}(\rho_{J})&=(N_J)^{-1}\sum_{k=0}^{\lfloor  \delta^{-s}\rfloor}p^{2J-k}(1-p)^{k}\frac{{2K\choose k}}{{2J\choose k}}|K,K-k\>\<K,K-k| +\mu_{J,K}.
\end{align*}
Using  Eq. (\ref{rho-decomp}) for  $\rho_{K}$, we bound the error as
\begin{align*}
\frac12\|\map{C}_{J\to K}(\rho_{J})-\rho_{K}\|_1 &= \frac12\left\|\sum_{k=0}^{\lfloor  \delta^{-s}\rfloor}(1-p)^k\left[\frac{p^{2J-k}}{N_J}\frac{{2K\choose k}}{{2J\choose k}}-\frac{p^{2K-k}}{N_{K}}\right]|K,K-k\>\<K,K-k|+\mu_{J,K}\right.\\
&\qquad\left.-\sum_{k=\lfloor  \delta^{-s}\rfloor+1}^{2K}\frac{p^{2K-k}(1-p)^k}{N_{K}}|K,K-k\>\<K,K-k|\right\|_1\\
&\le \frac12\sum_{k=0}^{\lfloor  \delta^{-s}\rfloor}(1-p)^kp^{-k}\left|\frac{p^{2J}}{N_J}\frac{{2K\choose k}}{{2J\choose k}}-\frac{p^{2K}}{N_{K}}\right|+\frac12\Tr[\mu_{J,K}]+\frac12\sum_{k=\lfloor  \delta^{-s}\rfloor+1}^{2K}\frac{p^{2K-k}(1-p)^k}{N_{K}}\\
&\le\frac{p}{4p-2}\max_{k\in[0,\lfloor  \delta^{-s}\rfloor]}\left|\frac{p^{2J}}{N_J}\frac{{2K\choose k}}{{2J\choose k}}-\frac{p^{2K}}{N_{K}}\right|+\frac12\Tr[\mu_{J,K}]+\left(\frac{1-p}{p}\right)^{\lfloor  \delta^{-s}\rfloor}
\end{align*}
Since $p>1/2$ and $s>0$, it is obvious that the third term in the last inequality vanishes exponentially in $J$, and we need only to show that the first term and the second term also vanish as $J$ grows.

For the first error term, we have the following expansion since $k\ll J$:
\begin{align*}
&\left|\frac{p^{2J}}{N_J}\frac{{2K\choose k}}{{2J\choose k}}-\frac{p^{2K}}{N_{K}}\right|\\
=&\frac{p^{2K}}{N_{K}}\left|p^{2J-2K}\frac{N_{K}}{N_{J}}\frac{{2K\choose k}}{{2J\choose k}}-1\right|\\
=&\frac{p^{2K}}{N_{K}}\left|\frac{1-\left(\frac{1-p}{p}\right)^{2K+1}}{1-\left(\frac{1-p}{p}\right)^{2J+1}}\frac{{2K\choose k}}{{2J\choose k}}-1\right|\\
=&\frac{p^{2K}}{N_{K}}\left|\frac{1-\left(\frac{1-p}{p}\right)^{2K+1}}{1-\left(\frac{1-p}{p}\right)^{2J+1}}e^{k\ln(K/J)+(2J-k+1)\ln[1-k/(2J)]-(2K-k+1)\ln[1-k/(2K)]+O(J^{-1})}-1\right|.
\end{align*}
Recalling that $\delta=(J-K)/J$, it is straightforward to verify that 
\begin{align*}
e^{k\ln(K/J)}&=1-k\delta+O(k\delta^{2})\\
e^{(2J-k+1)\ln[1-k/(2J)]-(2K-k+1)\ln[1-k/(2K)]}&=1-\frac{k^2  \delta}{4J}+O\left(k^3\delta J^{-2}\right).
\end{align*}
Substituting the above equations into the expression of the first error term, we have
\begin{align*}
\max_{k\in[0,\lfloor \delta^{-s}\rfloor]}\left|\frac{p^{2J}}{N_J}\frac{{2K\choose k}}{{2J\choose k}}-\frac{p^{2K}}{N_{K}}\right|&=\frac{p^{2K}}{N_{K}}\max_{k\in[0,\lfloor \delta^{-s}\rfloor]}\left|-k\delta-\frac{k^2  \delta}{4J}+O\left(k^3\delta J^{-2}\right)+O(J^{-1})\right|\\
&\le \frac{p^{2K}}{N_{K}}\left[\delta^{1-s} +O(J^{-1})\right]\\
&=\frac{2p-1}{p}\cdot \delta^{1-s} [1+O(\delta^s)].
\end{align*}

For the second term, we have
\begin{align*}
\Tr[\mu_{g,J,K}]&=1-(N_J)^{-1}\sum_{m=J-\lfloor \delta^{-s}\rfloor}^{J}p^{J+m}(1-p)^{J-m}\frac{{2K\choose J-m}}{{2J\choose J-m}}\\
&\le 1-(N_J)^{-1}\sum_{m=J-\lfloor \delta^{-s}\rfloor}^{J}p^{J+m}(1-p)^{J-m}\min_{m'\in[J-\lfloor \delta^{-s}\rfloor,J]}\frac{{2K\choose J-m'}}{{2J\choose J-m'}}\\
&\le 1-(N_J)^{-1}\sum_{m=J-\lfloor \delta^{-s}\rfloor}^{J}p^{J+m}(1-p)^{J-m}\\
&=1-\frac{1-\left(\frac{1-p}p\right)^{\lfloor \delta^{-s}\rfloor+1}}{1-\left(\frac{1-p}p\right)^{2J+1}}\\
&\le\left(\frac{1-p}p\right)^{\lfloor \delta^{-s}\rfloor+1},
\end{align*}
which vanishes exponentially fast as $J$ grows.  Finally, combining the above calculations, the error of the conversion can be bounded as
\begin{align}\label{e-bound-2}
\frac12\|\map{C}_{J\to K}(\rho_{J})-\rho_{K}\|_1  &\le\frac{\delta^{1-s}}2 [1+O(\delta^s)] \, .
\end{align} 
for any $s>0$.

\section{Precision analysis for known spectrum.}
The compression protocol for known spectrum is characterized by the couple $(\map{E},\map{D})$, where the encoding channel is
\begin{align*}
\map{E}(\rho)=\sum_{J=0}^{n/2} \map{C}_{J\to J_0}
\left[\Tr_{\spc{M}_J}\left(\Pi_J \rho \Pi_J \right) \right]
\end{align*}
where  $\Pi_J$ is the projector on $\spc R_J\otimes \spc M_J$ and $\Tr_{\spc M_{J}}$ is the partial trace over $\spc M_J$.  The decoding channel is
\begin{align*}
\map{D}(\sigma)=\bigoplus_{K=0}^{n/2} q_{K} \left[\map{C}_{J_0\to K}(\sigma)\otimes\frac{I_{m_{K}}}{m_{K}}\right].
\end{align*}

It is then straightforward to check that, when the input state is $\rho_{g}^{\otimes n}$, the output state of the protocol will be
\begin{align*}
\map{D}\circ\map{E}\left(\rho_g^{\otimes n}\right)=\bigoplus_{K}q_{K}\left[\sum_J q_J\left(\map{C}_{J_0\to K}\circ\map{C}_{J\to J_0}\right)(\rho_{g,J})\otimes \frac{I_{m_{K}}}{m_{K}}\right].
\end{align*}

Now we evaluate the performance of the protocol.
The error can be expressed and bounded as in the following.
\begin{align*}
\epsilon&=\max_{g\in\grp{SU}(2)}\frac12\left\|\map{D}\circ\map{E}\left(\rho_g^{\otimes n}\right)-\rho_g^{\otimes n}\right\|_1\\
&=\max_{g}\frac12\sum_{K}q_{K}\left\|\rho_{g,K}-\sum_J q_J\left(\map{C}_{J_0\to K}\circ\map{C}_{J\to J_0}\right)(\rho_{g,J})\right\|_1\\
&\le \frac12\sum_{J,K}q_J q_{K}\left\|\rho_{K}-\left(\map{C}_{J_0\to K}\circ\map{C}_{J\to J_0}\right)(\rho_{J})\right\|_1 \, ,
\end{align*}
having used the covariance of the universal cloning channel.  

Now, recall  that, for large $n$,   the distribution $\{q_J\}$ is peaked around $J_0$. Using this fact, we can define the set 
\begin{align}
\set{S}:=[J_0- \sqrt{n^{1+s}},J_0+ \sqrt{n^{1+s}}]
\end{align}
for some positive parameter $s$ to be specified later, so that $\lim_{n\to\infty}\sum_{J\not\in\set{S}}q_J=0$.  Then, we  continue bounding the error as
\begin{align}\nonumber
\epsilon&\le \frac12\sum_{J\not\in\set{S},K}q_{J}q_{K}+ \frac12\sum_{K\not\in\set{S},J}q_{J}q_{K}+\frac12\sum_{J,K\in\set{S}}q_J q_{K}\|\rho_{K}-\left(\map{C}_{J_0\to K}\circ\map{C}_{J\to J_0}\right)(\rho_{J})\|_1\\
\nonumber&= \sum_{J\not\in\set{S}}q_{J}+\frac12\sum_{J,K\in\set{S}}q_J q_{K}\|\rho_{K}-\left(\map{C}_{J_0\to K}\circ\map{C}_{J\to J_0}\right)(\rho_{J})\|_1\\
&\le\sum_{J\not\in\set{S}}q_{J}+\max_{J,K\in\set{S}}  \|\rho_{K}-\map{C}_{J\to K}(\rho_{J})\|_1 \, ,\label{e-bound}
\end{align}
where the last inequality comes from the bound 
\begin{align*}
\|\rho_{K}-\left(\map{C}_{J_0\to K}\circ\map{C}_{J\to J_0}\right)(\rho_{J})\|_1&\le \left\|\rho_{K}-\map{C}_{J_0\to K}(\rho_{J_0})\right\|_1+\left\|\map{C}_{J_0\to K}(\rho_{J_0})-\left(\map{C}_{J_0\to K}\circ\map{C}_{J\to J_0}\right)(\rho_{J})\right\|_1\\
&\le \|\rho_{K}-\map{C}_{J_0\to K}(\rho_{J_0})\|_1+\|\rho_{J_0}-\map{C}_{J\to J_0}(\rho_{J})\|_1\\
&\le 2\max_{J,K\in\set{S}}  \|\rho_{K}-\map{C}_{J\to K}(\rho_{J})\|_1.
\end{align*}

Now, we show that both terms in Eq. (\ref{e-bound}) vanish in the large $n$ limit.
To handle the first term, we use the explicit expression of $q_J$ \cite{yang-chiribella-2016-prl}, whose derivation is provided here for convenience of the reader: 
\begin{align*}
q_J&=\Tr\left[\Pi_J\rho_g^{\otimes n}\right]\\
&=m_J\sum_{m=-J}^{J}p^{n/2+m}(1-p)^{n/2-m}\\
&=\frac{(2J+1)[p^{n/2+J+1}(1-p)^{n/2-J}-p^{n/2-J}(1-p)^{n/2+J+1}]}{(2p-1)(n+1)}{n+1\choose n/2+J+1},
\end{align*}
having used the expression of the multiplicity $m_J=(2J+1){n+1\choose n/2+J+1}/(n+1)$. Rearranging the terms we get
\begin{align}
q_{J}=\frac{2J+1}{2J_0}&\left[  B\left(\frac n2 +  J+1 \right) -B\left(\frac n2 - J \right)\right]\label{qJ}
\end{align}
where $B(k)=p^{k}(1-p)^{n-k}{n\choose k}$ and $J_0 = (p-1/2)(n+1)$.

Using Eq. (\ref{qJ}), we then have
\begin{align}
\sum_{J\not\in\set{S}}q_J&=1-\sum_{J=J_0-\sqrt{n^{1+s}}}^{J_0+\sqrt{n^{1+s}}} \frac{2J+1}{2J_0}\left[   B\left(\frac n2 +  J+1 \right) -B\left(\frac n2 - J \right)\right]\nonumber \\
&\le1-\frac{2J_0-2\sqrt{n^{1+s}}+1}{2J_0}\sum_{J=J_0-\sqrt{n^{1+s}}}^{J_0+\sqrt{n^{1+s}}} B\left(\frac n2 +  J+1 \right) +\frac{2J_0+2\sqrt{n^{1+s}}+1}{2J_0}\sum_{J=J_0-\sqrt{n^{1+s}}}^{J_0+\sqrt{n^{1+s}}} B\left(\frac n2 - J \right)\nonumber\\
&\le 1-\frac{2J_0-2\sqrt{n^{1+s}}+1}{2J_0}\left[1-2\exp\left(-2n^{1+s}p^{-2}\right)\right]+2\exp\left[-2\left(\frac{1-p}{p}\right)^2n\right]\nonumber\\
&\le \frac{\sqrt{n^{1+s}}}{J_0}+2\exp\left(-\frac{2n^{1+s}}{p^2}\right)+2\exp\left[-2\left(\frac{1-p}{p}\right)^2n\right],\label{term1}
\end{align} 
having used the Hoeffding's inequality in the second last inequality. From the above inequalities, it is clear that for any positive threshold we can choose an $s$ small enough so that this term is bounded by the threshold for large enough $n$. 

 On the other hand, we notice that $J\approx K$ for any $J, K\in\set{S}$, and thus the second error term also vanishes.
Substituting $\delta\le (2\sqrt{n^{1+s}})/J_0$ into Eq. (\ref{lemma}), we get that
\begin{align}\label{term2}
\max_{J,K\in\set{S}}  \|\rho_{g,K}-\map{C}_{J\to K}(\rho_{g,J})\|_1\le n^{-\frac{1-s}{2}+s'}+O\left(n^{-\frac{1-s}{2}}\right)\qquad\forall s'>0.
\end{align}

Summarizing from Eq. (\ref{term1}) and Eq. (\ref{term2}), we have shown that $\epsilon\le O\left(n^{-\frac12+s}\right)$ for arbitrarily small $s>0$.

\section{Elementary properties of sufficient statistics}

Here we complete the argument given in the main text, showing that if $\set E'$ is a sufficient statistics for $\set E$, then \emph{i)}   $\set E$ and $\set E'$ have the same Holevo information and \emph{ii)} $\set E$ can be stored in a memory of $q$ qubits with error $\epsilon$ if and only if $\set E'$ can be stored in a memory of the same size, with the same error.   

By definition, the fact that  $\set{E}'$ is  a sufficient statistics means that there exist encoding and decoding channels $(\map E_0, \map D_0)$ such that  reversible map $\map{R}$ from any state $\rho_x\in\set{E}$ to the state $\rho_x'\in\set{E}'$, in formula
\begin{align}\label{sufficient}
  \map E_0 (\rho_x)  =  \rho_x'  \qquad {\rm and} \qquad  \map D_0  (\rho_x')  =  \rho_x  \, ,
\end{align}
for every possible $x$. 
Using the above relation, it is easy to show that every compression protocol for the ensemble $\set E$---say, $(\map E,\map D)$---can be turned into a compression protocol for the ensemble $\set E'$---call it  $(\map E',\map D')$---by defining  
\begin{align*}
\map E'    :  =         \map E \circ  \map D_0  \qquad {\rm and}   \qquad \map D'   :=      \map E_0\circ  \map D \, .
\end{align*}
Likewise, every compression protocol for $\set E'$---say $(\map E',\map D')$---can be turned into a compression protocol for $\set E$---call it $(\map E, \map D)$---by defining  
\begin{align}
\map E    :  =         \map E' \circ  \map E_0  \qquad {\rm and}   \qquad \map D   :=      \map D_0\circ  \map D \, .
\end{align}
Hence, the ensembles $\set E$ and $\set E'$ can be compressed in the same quantum memory with the same amount of error.

Moreover, Eqs. (\ref{sufficient}) and the monotonicity of Holevo's information imply the relations $\chi (\set E') \le \chi (\set E)$ and $\chi(\set E) \le \chi (\set E')$, whence $\chi(\set E')  \equiv \chi (\set E)$.   

\section{Optimality of the compression protocol for known spectrum.}
In this section we present the complete prove  for the optimality of our protocol for compressing qubit states with  known spectrum.  
 We choose the sufficient statistics $\set E'= \{\bigoplus_J q_J\rho_{g,J}, \d g\}$, which has effective dimension $d_{\set{E}'}=(n/2+1)^2$. Recall from the Letter the following bound
\begin{align}\label{bound00}
\log d_{\rm enc}\ge \chi\left(\set{E}\right)-4\epsilon\log n+4\epsilon-2\mu(\epsilon)+O(1).
\end{align}
Next, explicit calculation shows that the Holevo information of the ensemble $\set{E}$ can be expressed as
\begin{align}
\chi\left(\set{E}\right) &=-nH(\rho_g)+H(\{q_{J}\})+\sum_J q_J\left[\log (2J+1)+\log m_J\right]\label{bound-step1}
\end{align}
From a previous work [see Eqs. (7), (10) and (11) of \cite{hayashi-2010-cmp}], we know that
\begin{align}\label{bound-step2}
\sum_J q_J \left[\log(2J+1)+\log m_J\right]= \frac{1}{2}\log n+nH(\rho_g)+O(1).
\end{align}
For the entropy of the probability distribution $\{q_J\}$, we first notice that by definition [cf. Eq. (\ref{qJ})], the entropy of $\{q_J\}$ is
\begin{align}\label{inter1}
H\left(\{q_J\}\right)= -\sum_J q_J\left\{\log\frac{2J+1}{2J_0}+\log B\left(\frac n2+J+1\right)+\log\left[1-\frac{B\left(n/2-J\right)}{B(n/2+J+1)}\right]\right\}.
\end{align}

Next, we calculate the three terms in Eq. (\ref{inter1}) separately. Notice from Eq. (6) of \cite{hayashi-2010-cmp} that asymptotically the first term is
\begin{align}
-\sum_J q_J\log\frac{2J+1}{2J_0}&=\log(2J_0)-\sum_J q_J\log(2J+1)\nonumber\\
&=\log (n+1)+\log(2p-1)-\log(2p-1)-\log n+o(1)\nonumber\\
&=o(1)\label{inter2},
\end{align}
which vanishes with the growth of $n$. By explicit expanding the binomial distribution, the second term can be calculated as
\begin{align}
&-\sum_J q_J\log B\left(\frac n2+J+1\right)\nonumber\\
=&-\sum_J q_J\log\left[p^{p(n/2+J+1)}(1-p)^{(1-p)(n/2-J)}{n+1\choose \frac n2+J+1}\right]\nonumber\\
=&-p\log p\sum_J q_J\left(\frac n2+J+1\right)-(1-p)\log(1-p)\sum_J q_J\left(\frac n2-J\right)-\sum_J q_J\log {n+1\choose \frac n2+J+1}\nonumber\\
=&nH(\{p,1-p\})-\sum_J q_J \log {n\choose \frac n2+J}+O(1)\nonumber\\
=&nH(\{p,1-p\})-nH(\{p,1-p\})+\frac{1}{2}\log n+O(1)\nonumber\\
=&\frac{1}{2}\log n+O(1)\label{inter3},
\end{align}
having used Eq. (11) of \cite{hayashi-2010-cmp} in the second last step. 
Finally, the last term in Eq. (\ref{inter1}) can be evaluated as
\begin{align}
-\sum_J q_J\log\left[1-\frac{B\left(n/2-J\right)}{B(n/2+J+1)}\right]
&=-\sum_J q_J\log\left[1-\left(\frac{1-p}{p}\right)^{2J+1}\frac{n+2J+2}{n-2J}\right]\nonumber\\
&=O\left[\left(\frac{1-p}{p}\right)^{2J_0}\right]\nonumber\\
&=o(1).\label{inter4}
\end{align}
Substituting Eqs. (\ref{inter2}), (\ref{inter3}) and (\ref{inter4}) into Eq. (\ref{inter1}), we immediately get that
\begin{align}\label{bound-step3}
H(\{q_J\})=\frac12\log n+O(1).
\end{align}
Substituting Eqs. (\ref{bound-step1}), (\ref{bound-step2}) and (\ref{bound-step3}) into Eq. (\ref{bound00}), we bound the memory size as
\begin{align}\label{bound-final}
\log d_{\rm enc}(\set{E})&\ge\log n-4\epsilon\log n+4\epsilon-2\mu(\epsilon)+O(1).
\end{align}
When $\epsilon$ is vanishing, the leading order in the bound (\ref{bound-final}) is $\log n$. We thus conclude that our protocol for the known-spectrum compression is asymptotically optimal.

\section{Precision analysis for the full model compression}
Let us first recall the details of the compression protocol.  The protocol uses a  partition of the set $\{0,\dots,n/2\}$ into $t=O(\sqrt{n})$ intervals $\set L_1, \dots \set L_t$, defined as follows:
\begin{align*}
\set{L}_m&=\{  (m-1) \, \lfloor r\sqrt{n}\rfloor ,\dots, m \,  \lfloor r\sqrt{n}\rfloor  -1 \}  \, , \qquad m=  1,\dots,  t-1 \\
\set{L}_t&=\{n/2\} \, ,
\end{align*} 
 where $r$ is a parameter, chosen so that $\lfloor r\sqrt{n}\rfloor\times(t-1)=n/2$. We denote by $$\set{Med}=\{\lfloor r\sqrt{n}\rfloor/2, 3\lfloor r\sqrt{n}\rfloor/2, \dots\}$$ the collection of all medians of these subsets.
In the encoder, we measure the total spin using the POVM $\{\Pi_J\}_J$ and store the index $i(J)$ that $J\in\set{L}_{i(J)}$.
For convenience, we define a map $f$ which takes any $J\in\{0,\dots,n/2\}$ to the median of the subset containing $J$, formally defined as 
\begin{align*}
f: J\to J_{\rm med}\in\set{Med}\quad{\rm s.t.}\quad J_{\rm med}\in\set{L}_{i(J)}.
\end{align*}
Then the encoding channel can be represented as 
\begin{align*}
\map{E}(\rho):=
\sum_{J=0}^{n/2}\map{C}_{J\to f(J)}\left(\Tr_{\spc{M}_J}[\Pi_J \rho \Pi_J]\right)
\otimes |i(J)\>\<i(J)| \, .
\end{align*}
The  decoding channel is
\begin{align*}
\map{D}\left(\sum_{i} \sigma_{i}\otimes|i\rangle \langle i|\right):=\bigoplus_{K\in \set L_i}
\frac{1}{|\set{L}_i |}  \, \left[\map{C}_{f(K)\to K}(\sigma_{i})\otimes\frac{I_{m_{K}}}{m_{K}}\right]
\end{align*}
Note that pure states are compressed with zero error. Indeed,  when the state $\rho_g$ is pure ($p=1$ or $p=0$), the state $\rho_g^{\otimes n}$ is contained in the symmetric subspace, with $J  = n/2$.   By the definition of $\map E$ and $\map D$, we have 
\[   \map D \circ \map E (\rho_{n/2}) =  \rho_{n/2} \]    
for every state $\rho_{n/2}$ with support in the symmetric subspace. 

Let us focus now on the mixed state case ($0<p<1$). The output state of the protocol can be expressed as
\begin{align*}
(\map{D}\circ\map{E})(\rho_g^{\otimes n})=\bigoplus_{J=0}^{n/2}\left[\sum_{K\in\set{L}_{i(J)}}\frac{q_{K}}{|\set{L}_{i(J)} |}\map{C}_{f(J)\to J}\circ\map{C}_{K\to f(J)}(\rho_{g,K})\right]\otimes\frac{I_{m_J}}{m_J}.
\end{align*}
Noticing that the encoder and the decoder fare equally well on all input states, the error of the protocol can be written as
\begin{align*}
\epsilon&=\max_{g}\frac12\left\|(\map{D}\circ\map{E})(\rho_g^{\otimes n})-\rho_g^{\otimes n}\right\|_1\nonumber\\
&=\frac12\left\|\bigoplus_{J=0}^{n/2}\left[\sum_{K\in\set{L}_{i(J)}}\frac{q_{K}}{|\set{L}_{i(J)}|}\map{C}_{f(J)\to J}\circ\map{C}_{K\to f(J)}(\rho_{K})\right]\otimes\frac{I_{m_J}}{m_J}-\bigoplus_{J=0}^{n/2}q_J\left(\rho_{J}\otimes\frac{I_{m_J}}{m_J}\right)\right\|_1\nonumber\\
&=\frac12\sum_{J}\left\|\left[\sum_{K\in\set{L}_{i(J)}}\frac{q_{K}}{|\set{L}_{i(J)} |}\map{C}_{f(J)\to J}\circ\map{C}_{K\to f(J)}(\rho_{K})\right]-q_J \rho_{J}\right\|_1.
\end{align*}
To further bound the error, we shall use the concentration property of the distribution $\{q_J\}$. Explicitly, we define a set $\set{S}$ as
\begin{align*}
\set{S}=\left\{\lfloor J_0-cr\sqrt{n}\rfloor,\dots,\lfloor J_0+cr\sqrt{n}\rfloor\right\}
\end{align*}
with a parameter $c>0$ controlling $|\set{S}|$. For any $t>0$ we can choose $c$ to be large enough that $\lim_{n\to\infty}\sum_{J\not\in\set{S}}q_J=0$ as shown later. Separating the tail error term from the rest, we get that
\begin{align*}
\epsilon&=\frac12\sum_{J\in\set{S}}\left\|\left[\sum_{K\in\set{L}_{i(J)}}\frac{q_{K}}{|\set{L}_{i(J)} |}\map{C}_{f(J)\to J}\circ\map{C}_{K\to f(J)}(\rho_{K})\right]-q_J \rho_{J}\right\|_1+\frac12\sum_{J\not\in\set{S}}\left\|\left[\sum_{K\in\set{L}_{i(J)}}\frac{q_{K}}{|\set{L}_{i(J)}|}\map{C}_{f(J)\to J}\circ\map{C}_{K\to f(J)}(\rho_{K})\right]-q_J \rho_{J}\right\|_1.
\end{align*}
We further split the error within $\set{S}$ into two terms: the first error term is the imprecision of the adapter, while the second error term is the error of the interpolation. Precisely, we have:
\begin{align}
\epsilon&\le \epsilon_1+\epsilon_2+\epsilon_3\label{e-1}\\
\epsilon_1&=\frac12\sum_{J\in\set{S}}\left\|\left[\sum_{K\in\set{L}_{i(J)}}\frac{q_{K}}{\lfloor r\sqrt{n}\rfloor}\map{C}_{f(J)\to J}\circ\map{C}_{K\to f(J)}(\rho_{K})\right]-\left[\sum_{K\in\set{L}_{i(J)}}\frac{q_{K}}{\lfloor r\sqrt{n}\rfloor}\right]\rho_{J}\right\|_1\\
\epsilon_2&=\frac12\sum_{J\in\set{S}}\left\|\left[\sum_{K\in\set{L}_{i(J)}}\frac{q_{K}}{\lfloor r\sqrt{n}\rfloor}\right]\rho_{J}-q_J\rho_{J}\right\|_1\\
\epsilon_3&=\frac12\sum_{J\not\in\set{S}}\left\|\left[\sum_{K\in\set{L}_{i(J)}}\frac{q_{K}}{\lfloor r\sqrt{n}\rfloor}\map{C}_{f(J)\to J}\circ\map{C}_{K\to f(J)}(\rho_{K})\right]-q_J\rho_{J}\right\|_1.
\end{align}
Now, we show the details of bounding each of these three error terms respectively. First, the error term $\epsilon_1$, namely the imprecision of the adapter, can be upper bounded as
\begin{align}
\epsilon_1&\le\frac12\sum_{J\in\set{S}}\left\|\left[\sum_{K\in\set{L}_{i(J)}}\frac{q_{K}}{\lfloor r\sqrt{n}\rfloor}\map{C}_{f(J)\to J}\circ\map{C}_{K\to f(J)}(\rho_{K})\right]-\left[\sum_{K\in\set{L}_{i(J)}}\frac{q_{K}}{\lfloor r\sqrt{n}\rfloor}\right]\map{C}_{f(J)\to J}\left(\rho_{f(J)}\right)\right\|_1\nonumber\\
&\qquad+\frac12\sum_{J\in\set{S}}\left\|\left[\sum_{K\in\set{L}_{i(J)}}\frac{q_{K}}{\lfloor r\sqrt{n}\rfloor}\right]\map{C}_{f(J)\to J}\left(\rho_{f(J)}\right)-\left[\sum_{K\in\set{L}_{i(J)}}\frac{q_{K}}{\lfloor r\sqrt{n}\rfloor}\right]\rho_{J}\right\|_1\nonumber\\
&\le\left[\sum_{J=0}^{n/2}\sum_{K\in\set{L}_{i(J)}}\frac{q_{K}}{\lfloor r\sqrt{n}\rfloor}\right]\frac12\left\{\max_{J\in\set{S}}\left\|\map{C}_{f(J)\to J}\left(\rho_{f(J)}\right)-\rho_{J}\right\|_1+\max_{J\in\set{S}}\left\|\map{C}_{J\to f(J)}\left(\rho_{J}\right)-\rho_{f(J)}\right\|_1\right\}\nonumber\\
&=\frac12\left\{\max_{J\in\set{S}}\left\|\map{C}_{f(J)\to J}\left(\rho_{f(J)}\right)-\rho_{J}\right\|_1+\max_{J\in\set{S}}\left\|\map{C}_{J\to f(J)}\left(\rho_{J}\right)-\rho_{f(J)}\right\|_1\right\}\nonumber\\
&\le\max_{J\in\set{S}}\max_{K\in\set{L}_{i(J)}}\left\|\map{C}_{J\to K}\left(\rho_{J}\right)-\rho_{K}\right\|_1\nonumber\\
&\le \left(\frac{r}{\sqrt{n}}\right)^{1-s}+O\left(\frac{r}{\sqrt{n}}\right)\qquad\forall s>0,\label{e-term1}
\end{align}
having used Eq. (\ref{lemma}) in the last step. Second, the error term $\epsilon_2$, namely the error of the interpolation, can be upper bounded as
\begin{align}
\epsilon_2&=\frac12\sum_{J\in\set{S}}\left|\left[\sum_{K\in\set{L}_{i(J)}}\frac{q_{K}}{\lfloor r\sqrt{n}\rfloor}\right]-q_J\right|\nonumber\\
&\le \left(\frac12\sum_{J\in\set{S}}q_J\right)\max_{J\in\set{S}} \max_{K\in\set{L}_{i(J)}}\left|\frac{q_J}{q_{K}}-1\right|\nonumber\\
&\le \frac12\max_{J\in\set{S}} \max_{K\in\set{L}_{i(J)}}\left|\frac{q_K}{q_{J}}-1\right|.\label{step-e2-1}
\end{align}
Now, by Eq. (\ref{qJ}) we have
\begin{align*}
\frac{q_K}{q_{J}}&=\frac{2K+1}{2J+1}\cdot\frac{B\left(\frac{n}2+K+1\right)-B\left(\frac{n}2-K\right)}{B\left(\frac{n}2+J+1\right)-B\left(\frac{n}2-J\right)}.
\end{align*}
We further notice that, by the De Moivre-Laplace theorem, the binomial $B(k)$ can be approximated by a Gaussian for $J\in\set{S}$ and for large $n$. Precisely we have
$$B\left(\frac{n}2+J+1\right)=\frac{1}{\sqrt{2\pi np(1-p)}}\exp\left[-\frac{(J-J_0)^2}{2np(1-p)}\right]\left[1+O\left(\frac{1}{\sqrt{n}}\right)\right].$$
Moreover, noticing that the term $B\left(\frac{n}2-J\right)$ is exponentially small compared to $B\left(\frac{n}2+J+1\right)$, we have
\begin{align}\label{step-e2-21}
 \frac{q_K}{q_{J}}&\ge\frac{J_0-(c+1)r\sqrt{n}}{J_0-cr\sqrt{n}}\left\{1-\frac{cr^2}{p(1-p)}+O(c^2r^4)+O\left(\frac{1}{\sqrt{n}}\right)\right\}\\
  \frac{q_K}{q_{J}}&\le\frac{J_0-cr\sqrt{n}}{J_0-(c+1)r\sqrt{n}}\left\{1+\frac{cr^2}{p(1-p)}+O(c^2r^4)+O\left(\frac{1}{\sqrt{n}}\right)\right\}\label{step-e2-22}
\end{align}
Substituting (\ref{step-e2-21}) and (\ref{step-e2-22}) into (\ref{step-e2-1}), we have
\begin{align}\label{e-term2}
\epsilon_2\le\frac{cr^2}{2p(1-p)}+O\left(\frac{r}{\sqrt{n}}\right)+O(c^2r^4).
\end{align}
At last, the error term $\epsilon_3$, namely the tail term, can be upper bounded as
\begin{align}
\epsilon_3&\le\frac12\left(\sum_{J\not\in\set{S}}\sum_{K\in\set{L}_{i(J)}}\frac{q_{K}}{\lfloor r\sqrt{n}\rfloor}+\sum_{J\not\in\set{S}}q_J\right)\nonumber\\
&\le 1-\sum_{J=J_0-(c-1)r\sqrt{n}}^{J_0+(c-1)r\sqrt{n}}q_J\nonumber\\
&\le 2\exp\left[-\frac{2(c-1)^2r^2}{p^2}\right].\label{e-term3}
\end{align}
Finally, substituting Eqs. (\ref{e-term1}), (\ref{e-term2}) and (\ref{e-term3}) into (\ref{e-1}), we have
\begin{align}
\epsilon\le \left(\frac{r}{\sqrt{n}}\right)^{1-s}+\frac{cr^2}{2p(1-p)}+2\exp\left[-\frac{2(c-1)^2r^2}{p^2}\right]+O\left(\frac{r}{\sqrt{n}}\right)+O(c^2r^4)\qquad\forall s>0.
\end{align}
To ensure that the error can be bounded arbitrarily from above for small enough $r$ and big enough $n$, we can choose $c=r^{-1-\delta}$ for a small constant $\delta>0$. In this case the error bound reduces to
\begin{align*}
\epsilon\le \left(\frac{r}{\sqrt{n}}\right)^{1-s}+\frac{r^{1-\delta}}{2p(1-p)}+2\exp\left[-\frac{2}{r^{2\delta}p^2}\right]+O\left(\frac{r}{\sqrt{n}}\right)+O(r^{2-2\delta})\qquad\forall s>0.
\end{align*}
Recall that we are dealing with the mixed state case where $1/2<p<1$. We can choose, for instance, $r=1/(\log n)$ to make the above error bound to be vanishing with $n$. Conclusively, we have shown that for any state $\rho_g$ and any error threshold $\epsilon>0$ there exists suitable choice of $r$ and $n_0$ so that the error of the compression is smaller than $\epsilon$ for $n>n_0$.

\section{Optimality for the full-model compression.}
In this section, we prove that the full-model protocol is optimal when no prior information on the qubit state is available. A protocol for full-model should have vanishing error  fon any possible input ensemble of $n$ identically prepared qubit states. In particular, it should have vanishing error  on the ensemble \cite{hayashi-2010-cmp} $$\set{U}=\{\rho^{\otimes n}, \d g\,f(p)\d p\}  \, ,$$ where $f(p)$ is the probability distribution given by  
\[ f(p):=e^{c(p)}/\int_0^{1}\d p' e^{c(p')} \, , \qquad c(p):=2\log(2p-1)-[(4p-1)\log p+(4p-3)\log (1-p)]/(4p-2) \, .\]
 Explicitly, we show that every protocol  that compresses $\set{U}$ with vanishing error  requires a total memory size of  at least $(3/2)\log n$ qubits. 
 
As in the known-spectrum case discussed in the main text, we use the bound 
\begin{align}\label{app-bound0}
\log d_{\rm enc}(\set{U})\ge\chi\left(\set{U}\right)-2[\epsilon\log d_{\set{U}^{\min}}+\mu(\epsilon)]
\end{align}
where   $d_{\set U}^{\min}$ is the minimum of the effective dimension $d_{\set U'}$ over all ensembles $\set U'$ that are sufficient statistics for $\set U$.       We pick    the sufficient statistics $\set{U}'$  defined by  
\begin{align*}
\set{U}'=\left\{\bigoplus_J q_J\rho_{g,J}, \d g\,\d p\right\} \, .
\end{align*}
The effective dimension of the ensemble $\set U'$   is $d_{\set{U}'}=(n/2+1)^2$. Now, Theorem 1 of \cite{hayashi-2010-cmp} states that 
\begin{align}
\chi\left(\set{U}\right)=\frac{3}{2}\log n+O(1).\label{app-bound1}
\end{align}
Combining Eq. (\ref{app-bound0}) with Eq. (\ref{app-bound1}), we achieve the following lower bound on the memory size: 
\begin{align}\label{basta}
\log d_{\rm enc}(\set{U})\ge \frac32\log n -4\epsilon\log n+4\epsilon-\mu(\epsilon)+O(1).
\end{align}
For large $n$ and vanishing $\epsilon$, the leading order of the above bound is $(3/2)\log n$, as stated in the main text.  

Eq. (\ref{basta}) states that, \emph{if} a protocol uses a fully quantum memory, the minimum amount of qubits needed to compress a completely unknown state is $3/2 \log n$.  Since the quantum memory is a stronger resource than the classical memory, this result implies that the every protocol using $q$ qubits and $c$ classical bits to compress $n$ copies with vanishing error must satisfy the bound  $q+c  \ge 3/2 \log n$.    Our protocol saturates the bound, as it uses  $\log n$ qubits and $1/2 \log n$ bits.  A natural question is whether the number of qubits in our protocol can be further reduced.  The answer is negative, due to the following argument:  A compression protocol for the full model should also compress with vanishing error the ensemble $\set P  =  \{   \phi_g^{\otimes n} \,  ,  \d g\}$, where $\phi_g$ is the generic pure state $\phi_g=   g  \,  |0\>\<0|  \, g^\dag$.    In order to compress the ensemble $\set P$, one needs a memory of $\log n$ qubits \cite{yang-chiribella-2016-prl}.    Hence, our compression protocol uses \emph{i)} the minimum amount of qubits, and \emph{ii)}  the minimum total amount of qubits and classical bits.

\end{widetext}
\end{appendix}


\begin{thebibliography}{36}
\bibitem [{foo()}]{footnote}%
  \BibitemOpen
  \href@noop {} {}\bibinfo {note} {For concreteness, here we assume $n$ to be
  even and $J$ to be integer, but all the arguments hold also for odd $n$ and
  semi-integer $J$.}\BibitemShut {Stop}%
\end{thebibliography}
\end{document}